\documentclass[prl,aps,twocolumn,english,showkeys,aps,unsortedaddress,superscriptaddress,nofootinbib]{revtex4-2}
\usepackage{amsmath}
\usepackage{amssymb}
\usepackage{epsfig}
\usepackage{graphicx}
\usepackage[colorlinks]{hyperref}
\usepackage[T1]{fontenc}
\usepackage[utf8]{inputenc}
\usepackage{float}
\usepackage{color}
\usepackage[normalem]{ulem}
\usepackage{tikz}
\usetikzlibrary{positioning}
\usepackage{appendix}

\begin{document}

\title{Cluster Dynamics Stay Fast--Until Tricriticality}

\author{Minjun Jeon$^*$}
\affiliation{Department of Physics and Photon Science, Gwangju Institute of Science and Technology, Gwangju 61005, Republic of Korea}
\affiliation{Department of Physics and Astronomy, Seoul National University, Seoul 08826, Republic of Korea}

\author{Alexandros Vasilopoulos$^*$}
\affiliation{School of Mathematics, Statistics and Actuarial Science, University of Essex, Colchester CO4 3SQ, United Kingdom}

\author{Dong-Hee Kim} \email{dongheekim@gist.ac.kr}
\affiliation{Department of Physics and Photon Science, Gwangju Institute of Science and Technology, Gwangju 61005, Republic of Korea}

\author{V\'{i}ctor Mart\'{i}n-Mayor}
\affiliation{Departamento de F\'isica
T\'eorica, Universidad Complutense, 28040 Madrid, Spain}

\author{Nikolaos G. Fytas}\email{nikolaos.fytas@essex.ac.uk}
\affiliation{School of Mathematics, Statistics and Actuarial Science, University of Essex, Colchester CO4 3SQ, United Kingdom}

\def\thefootnote{*}

\date{\today}

\begin{abstract}
Cluster Monte Carlo algorithms are widely regarded as the most effective route to overcoming critical slowing down in lattice spin systems. Whether this acceleration persists in the presence of vacancies and multicritical fluctuations, however, remains unresolved. We address this question through a systematic dynamic-scaling study of hybrid cluster-local update schemes in the two-dimensional Blume-Capel model, which exhibits a line of continuous Ising-like transitions terminating at a tricritical point. Along the entire critical line, hybrid dynamics retain the near-optimal efficiency of pure cluster updates despite the presence of annealed vacancies. Strikingly, this acceleration collapses precisely at tricriticality, where the dynamic critical exponent reverts to the local-update value. We trace this breakdown to the correlated percolation of vacancies, whose emergent system-spanning geometry obstructs nonlocal relaxation in the spin sector. Our results identify a fundamental geometric limitation of cluster acceleration at tricriticality and establish vacancy percolation as the mechanism controlling dynamic universality in hybrid Monte Carlo dynamics.
\end{abstract}

\maketitle

\footnotetext{These authors contributed equally to this work.}

\textit{Introduction.}---Critical slowing down severely limits Monte Carlo simulations near continuous phase transitions, where diverging correlation lengths induce long-lived correlations between successive configurations. For local-update dynamics, the associated autocorrelation times scale with the linear system size as $\tau \sim L^{z}$, with the dynamic critical exponent $z$ taking the well-established value $z \approx 2.17$ for models in the two-dimensional Ising universality class under Metropolis dynamics~\cite{nightingale96,liu23,bisson25}. A major breakthrough came with the Fortuin-Kasteleyn random-cluster representation~\cite{FK72a,FK72b,FK72c}, which enabled collective-update algorithms such as those of Swendsen-Wang~\cite{swendsen87} and Wolff~\cite{wolff89}. By decorrelating critical fluctuations through nonlocal cluster flips, these methods dramatically suppress critical slowing down, often yielding nearly vanishing dynamic exponents. Nevertheless, this acceleration is not unrestricted: Li and Sokal established a rigorous lower bound relating cluster-dynamics autocorrelation times to the specific heat, thereby identifying energy-like fluctuations associated with cluster interfaces as an intrinsic slow mode~\cite{li:89}. Whether additional slow modes can emerge and dominate the dynamics in more complex settings remains an open question. This issue is particularly acute in systems with vacancies or annealed dilution, where the additional degrees of freedom are not naturally incorporated into standard cluster constructions, and becomes especially pressing near multicritical points, where distinct fluctuation channels may become simultaneously critical.

\begin{figure}[ht!]
    \centering
    \includegraphics[width=1.0\linewidth]{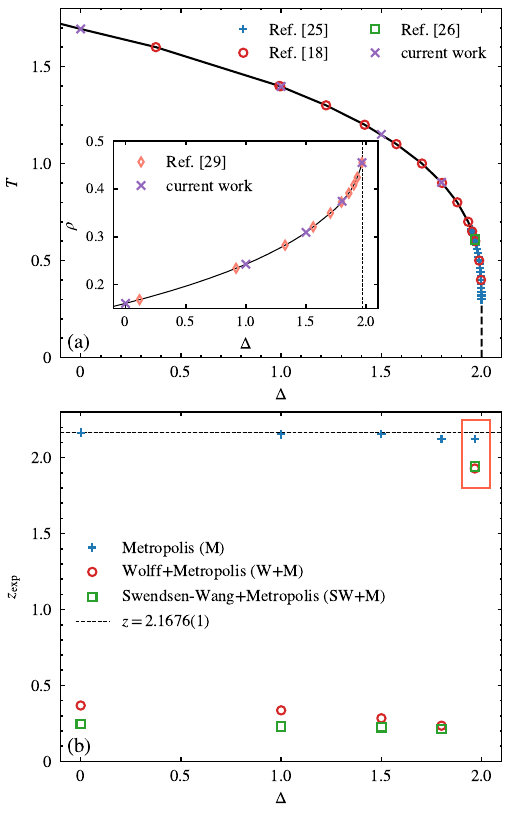}
    \caption{(a) Phase diagram of the square-lattice Blume-Capel model in the $(\Delta,T)$ plane, showing the continuous Ising-like transition line terminating at the tricritical point (green square), beyond which the transition becomes first order. The inset shows the average vacancy density $\rho$ along the continuous transition line and at tricriticality, combining data from Ref.~\cite{deng05} and the present work. The solid curve is a spline interpolation serving as a guide to the eye, and the dashed vertical line marks the tricritical-point location. (b) Dynamic critical exponent $z_{\rm exp}$, extracted from magnetization autocorrelations, along the transition line for the three update schemes indicated: Metropolis (M), Wolff$+$Metropolis (W$+$M), and Swendsen-Wang$+$Metropolis (SW$+$M). The horizontal dashed line marks the most accurate available Metropolis estimate for the two-dimensional Ising universality class, $z = 2.1676(1)$~\cite{bisson25}, while the red rectangle highlights the tricritical-point estimates.}
    \label{fig:phd_zexp}
\end{figure}

The spin-$1$ Blume-Capel model~\cite{blume66,capel66a,capel66b,capel67}, the minimal extension of the Ising universality class to include vacancies and a model of growing interest~\cite{wilding96,silva06,fytas-selke13,zierenberg15,zierenberg17,toldin17,hasenbusch20,moueddene24,mataragkas25a,mataragkas25b}, provides an ideal setting to address this problem. Here, we consider the square-lattice Blume-Capel model defined via
\begin{equation}
\mathcal{H}=-J\sum_{\langle \textbf{x}\textbf{y}\rangle}\sigma_{\textbf{x}}\sigma_{\textbf{y}}+\Delta\sum_{\textbf{x}}\sigma_{\textbf{x}}^2 = E_{J}+\Delta E_{\Delta},
\label{eq:hamiltonian}
\end{equation}
where the spins $\sigma_\textbf{x}\in\{-1,0,+1\}$ are located in the nodes
of an $N = L \times L$ square lattice with periodic boundary conditions, $\langle \textbf{x}\textbf{y}\rangle$ indicates summation over nearest neighbors, and $J > 0$ is the ferromagnetic exchange coupling. The parameter $\Delta$
is known as the crystal-field coupling that controls the density of vacancies ($\sigma_{\textbf{x}} = 0$). For $\Delta \rightarrow -\infty$ vacancies are suppressed and the model becomes equivalent to the simple Ising ($\sigma_{\textbf{x}} = \pm 1$) ferromagnet. The phase diagram of the Blume-Capel model contains a line of continuous Ising-like transitions terminating at a tricritical point $(\Delta_{\rm tp}, T_{\rm tp})=[1.96582(1), 0.60858(5)]$~\cite{qian05,kwak15,jung17,blote19}, beyond which the transition becomes first order~\cite{kwak15,jung17,mataragkas25b}; see also Fig.~\ref{fig:phd_zexp}(a) for a sketch of the phase boundary. Hybrid Monte Carlo schemes combining cluster updates for the $\pm 1$ spins with local updates for vacancies have proven highly effective along the critical line~\cite{zierenberg17,fytas18}. Yet their dynamic efficiency has never been systematically characterized, and earlier work suggested a possible breakdown near tricriticality~\cite{deng05}.

In this Letter, we present a systematic dynamic-scaling study of hybrid cluster-local algorithms along the continuous transition line and at the tricritical point of the two-dimensional Blume-Capel model. By combining autocorrelation analysis with geometric diagnostics of cluster connectivity, we examine how annealed vacancies influence the relaxation spectrum and the efficiency of nonlocal updates near multicriticality. This approach reveals that the performance of hybrid cluster dynamics is governed not solely by the critical spin fluctuations targeted by collective updates, but also by the geometric organization of vacancy degrees of freedom. Our findings establish a direct link between dynamic universality and correlated geometric fluctuations, providing a broader framework for understanding the limitations of cluster acceleration in diluted and multicritical systems.

\textit{Simulations and observables.}---We study both local Metropolis dynamics and hybrid cluster-local schemes in the grand-canonical ensemble. In the hybrid implementations, the $\pm 1$ sector is updated through Wolff~\cite{wolff89} or Swendsen-Wang~\cite{swendsen87} cluster moves, followed by a full-lattice Metropolis sweep over all spins, including vacancies. Several update mixtures were examined and found to yield statistically consistent dynamic behavior at tricriticality. Dynamic measurements and cluster-percolation analyses were performed for system sizes up to $L = 512$, with statistical uncertainties estimated via jackknife resampling. Critical inverse temperatures $\beta_{\rm c}(\Delta)$ (with $\beta\equiv1/T$) along the continuous transition line were determined from finite-size scaling analyses of Binder-cumulant and correlation-length crossings for systems up to $L = 4096$; the corresponding scaling analysis is presented in the End Matter.

The efficiency of each algorithm is characterized through the normalized autocorrelation function of an observable $A$,
\begin{equation}
C_A(t)=\frac{\langle A(t)A(0)\rangle-\langle A\rangle^2}
{\langle A^2\rangle-\langle A\rangle^2},
\label{eq:autocorrelation}
\end{equation}
where $A$ denotes the energy $E$, the absolute magnetization per spin $|m|=N^{-1}\left|\sum_{\mathbf{x}}\sigma_{\mathbf{x}}\right|$, or the vacancy density $\rho=1-e_{\Delta}$, with $e_{\Delta}=E_{\Delta}/N$. All three observables yield consistent dynamic scaling behavior. The exponential autocorrelation time $\tau_{\rm exp}$, associated with the slowest relaxation mode, is extracted from the asymptotic decay
$C_A(t)\sim \exp(-t/\tau_{\rm exp})$. The integrated autocorrelation time,
\begin{equation}
\tau_{\rm int}=\frac{1}{2}+\sum_{t=1}^{\infty}C_A(t),
\label{eq:tauint}
\end{equation}
quantifies the decorrelation time of the Markov chain and is estimated using the standard self-consistent windowing procedure~\cite{janke08}. Dynamic critical exponents are obtained from the finite-size scaling relation $\tau\sim L^z$. In the following, we report results for magnetization autocorrelations. Corresponding estimates obtained from energy and vacancy-density autocorrelations are fully consistent and are omitted for brevity. To ensure meaningful comparisons across update protocols, Monte Carlo time is normalized according to the average number of updated degrees of freedom; details are provided in the End Matter.

\begin{figure}[ht!]
    \centering
    \includegraphics[width=1.0\linewidth]{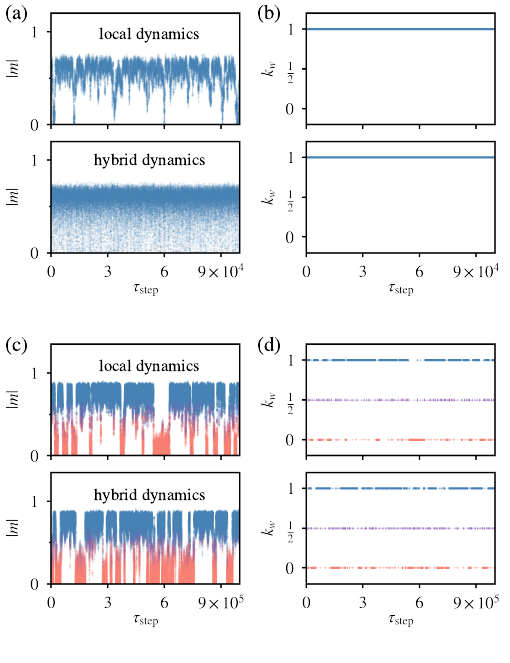}
    \caption{Monte Carlo dynamics and percolation signatures for $L=64$. Panels (a) and (b) correspond to $\Delta = 0$, while panels (c) and (d) to  the tricritical point $\Delta = \Delta_{\rm tp}$. The left column shows the time evolution of the absolute magnetization as a function of the Monte Carlo step count $\tau_{\rm step}$ for local (Metropolis) and hybrid (cluster-local) dynamics (see End Matter for the precise time normalization). Blue shading indicates the presence of wrapping clusters of $\pm 1$ spins, while salmon shading denotes wrapping vacancy clusters. Note the absence of vacancy percolation in panel (a). Each panel in the right column corresponds to its counterpart on the left and indicates the instantaneous percolation state: $k_w=1$ for wrapping clusters of $\sigma_{\textbf{x}} = \pm1$, $k_w=0$ for wrapping vacancy clusters ($\sigma_{\textbf{x}} = 0$), and $k_w=1/2$ for simultaneous wrapping of both vacancy and $\pm 1$ clusters.}
    \label{fig:mc_dyn_wc}
\end{figure}

Lastly, to probe the mechanism underlying the dynamic crossover, we analyze vacancy-cluster percolation using breadth-first-search identification of wrapping clusters on the torus and measure the corresponding wrapping probability $P_{\rm wrap}$~\cite{stauffer_introduction_1994,akritidis23}.

\textit{Results and discussion.}---Our central result is summarized in Fig.~\ref{fig:phd_zexp}. In particular, Fig.~\ref{fig:phd_zexp}(a) shows the phase diagram of the square-lattice Blume-Capel model, combining literature estimates with the high-precision critical-point determinations obtained in the present work along the continuous transition line from $\Delta=0$ to $\Delta=1.8$ (see Fig.~\ref{fig:beta_c} and Table~\ref{tab:beta_c} in the End Matter). The inset displays the average vacancy density $\rho$ along the critical line and at tricriticality. As expected, $\rho$ increases monotonically with $\Delta$, signaling the progressively enhanced role of vacancies as the tricritical point is approached. As shown below, this increase is accompanied by qualitative changes in the geometric organization of vacancies.

The corresponding dynamic critical exponents are shown in Fig.~\ref{fig:phd_zexp}(b) (see also Table~\ref{tab:zexp} in the End Matter). For local Metropolis dynamics, the estimates remain consistent with the established two-dimensional Ising and Blume-Capel value~\cite{nightingale96,liu23,bisson25}, with small deviations near tricriticality attributable to finite-size effects and the increased sensitivity to the precise location of the tricritical point. By contrast, the hybrid cluster-local schemes (Wolff$+$Metropolis and Swendsen-Wang$+$Metropolis) exhibit a strikingly different behavior. Along the continuous transition line, the dynamic exponent exhibits only a weak dependence on $\Delta$ and remains nearly an order of magnitude smaller than its Metropolis counterpart, demonstrating that annealed vacancies do not substantially impair the efficiency of nonlocal cluster relaxation.
This picture changes abruptly at tricriticality. All hybrid update schemes display a sharp increase in $z_{\rm exp}$, with values approaching the ones characterizing local Metropolis dynamics. The effect is robust across all algorithmic implementations considered, including different mixtures of Wolff, Swendsen-Wang, and Metropolis updates, indicating that increasing the number of cluster steps does not mitigate tricritical slowing down. This provides direct quantitative confirmation of the breakdown of cluster acceleration hypothesized in Ref.~\cite{deng05}, and establishes that the loss of efficiency is an intrinsic feature of tricritical dynamics rather than an artifact of a particular update protocol. 

The abrupt loss of cluster acceleration at tricriticality reflects the emergence of a geometric mechanism absent along the ordinary critical line: correlated vacancy percolation. This effect is illustrated in Fig.~\ref{fig:mc_dyn_wc}, which tracks the time evolution of the magnetization together with the instantaneous percolation state of the system. At $\Delta=0$ deep inside the second-order transition regime [panels (a) and (b)], vacancy percolation is entirely absent, while only the usual spin-cluster percolation persists. The same qualitative behavior is found throughout the continuous transition line for all $\Delta < \Delta_{\rm tp}$. By contrast, at tricriticality [panels (c) and (d)], both wrapping clusters of $\pm 1$ spins and system-spanning vacancy clusters are observed under both local and hybrid dynamics. The distinction is crucial. While the $\pm 1$ sector implies the Fortuin-Kasteleyn connectivity, vacancy sites do not admit such a bond representation since they do not interact directly [Eq.~\eqref{eq:hamiltonian}]. Their organization is therefore purely geometric. Earlier canonical-ensemble studies identified the corresponding vacancy-percolation threshold through bond activation~\cite{deng05}; here we show that in the grand-canonical ensemble the onset of geometric vacancy percolation coincides precisely with tricriticality. This establishes that the dynamic crossover observed in Fig.~\ref{fig:phd_zexp}(b) is tied to the appearance of a system-spanning correlated vacancy network. 

This interpretation is strongly supported by the behavior of the ratio $\tau_{\rm int}/\tau_{\rm exp}$ (Table~\ref{tab:tau}), which provides direct information on the spectral structure of the Markov dynamics~\cite{sokal97}. This ratio can be viewed as a Rayleigh-Ritz variational estimator of the dynamics' slowest mode: the closer $\tau_{\rm int}/\tau_{\rm exp}$ is to unity, the larger the spectral weight carried by the corresponding observable in the slowest relaxation channel. Table~\ref{tab:tau} shows that this occurs at tricriticality, and only at tricriticality, indicating that $|m|$ becomes an excellent variational probe of the dominant slow mode for the hybrid dynamics. Moreover, at tricriticality [Figs.~\ref{fig:mc_dyn_wc}(c) and (d)], the instantaneous value of $|m|$ directly tracks the percolation state of the system, demonstrating that its fluctuations are governed by fluctuations of the vacancy-percolation backbone itself. From this perspective, the evolution of $\tau_{\rm int}/\tau_{\rm exp}$ upon approaching tricriticality signals a crossover in the underlying relaxation mechanism: whereas along the critical line the dynamics remain constrained by the usual energy-like mode associated with the Li-Sokal bound~\cite{li:89}, at tricriticality this mode is supplanted by critical vacancy percolation. Once such a correlated vacancy backbone forms, collective updates of the $\pm1$ sector can no longer efficiently decorrelate the slowest long-wavelength fluctuations, causing the hybrid dynamics to revert to local-update scaling.

\begin{table}[]
\caption{Ratio $\tau_{\rm int}/\tau_{\rm exp}$ along the continuous transition line of the square-lattice Blume-Capel model, including the tricritical point at $\Delta_{\rm tp}=1.96582$, for the update schemes shown in Fig.~\ref{fig:phd_zexp}(b). Data are shown for $L=128$ for Metropolis dynamics and $L=512$ for the hybrid schemes, except at tricriticality where $L=128$ is used for all protocols.}
\begin{tabular*}{0.99\columnwidth}{@{\extracolsep{\fill}} lccc @{}}
\hline\hline
$\Delta$ &  & $\tau_{\rm int} / \tau_{\rm exp}$ &      \\
\hline
& M & W$+$M & SW$+$M\\
\hline
$0$   & $0.883(3)$ & $0.5191(8)$  & $0.5926(7)$ \\
\hline
$1$   & $0.919(3)$  & $0.4936(8)$ &  $0.5563(8)$   \\
\hline
$1.5$ & $0.875(4)$ & $0.4726(5)$  & $0.5034(9)$\\
\hline
$1.8$ & $0.898(4)$ &  $0.4396(8)$ & $0.4563(7)$ \\ 
\hline \hline
\multicolumn{4}{c}{\textbf{Tricritical Point}}      \\
 \hline
$1.96582$ & $0.982(7)$  & $0.968(6)$ & $0.973(6)$ \\
\hline \hline
\end{tabular*}
\label{tab:tau}
\end{table}

The vacancy density shown in the inset of Fig.~\ref{fig:phd_zexp}(a), combining results from Ref.~\cite{deng05} and the present work, increases monotonically with $\Delta$ and attains the value $\rho(\Delta_{\rm tp}) = 0.455(5)$ at the tricritical point. Since $\rho(\Delta < \Delta_{\rm tp}) < \rho(\Delta_{\rm tp})$, the system approaches a regime where the vacancy density is close to, but below, the uncorrelated percolation threshold $\rho_{\rm c} \simeq 0.5927$~\cite{ziff00}. 
\begin{figure}[ht!]
    \centering
    \includegraphics[width=1.0\linewidth]{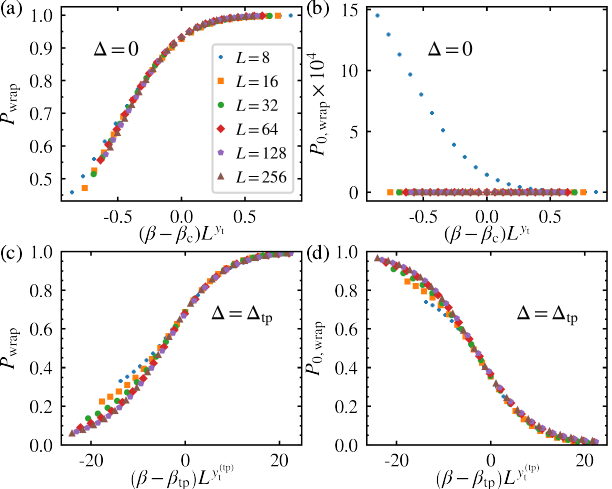}
    \caption{Finite-size scaling collapse of the wrapping probabilities of Fortuin-Kasteleyn clusters of $\pm1$ spins ($P_{\rm wrap}$, left column) and geometric vacancy clusters ($P_{0,\rm wrap}$, right column). Data are shown as functions of the scaling variable $(\beta-\beta_{\rm c})L^{y_{\rm t}}$ for $\Delta = 0$ and $\Delta = \Delta_{\rm tp}$, using $y_{\rm t} = 1$ along the continuous Ising-like transition line and the tricritical thermal exponent $y_{\rm t}^{\rm (tp)} = 9/5$ at $\Delta_{\rm tp}$~\cite{mataragkas25a}. The collapse of the vacancy-sector data at tricriticality demonstrates scale-invariant correlated vacancy percolation, while for $\Delta < \Delta_{\rm tp}$ vacancy clusters remain noncritical up to finite-size effects.}  
    \label{fig:pwrap_collapse}
\end{figure}
This suggests that vacancy percolation at tricriticality is not purely random but governed by strong spatial correlations. Previous work~\cite{deng05} showed that vacancy clusters defined via bond activation exhibit critical properties analogous to Fortuin-Kasteleyn clusters of the $\pm1$ spins at tricriticality. We substantiate this picture through a finite-size scaling analysis of the wrapping probabilities of both cluster sectors, shown in Fig.~\ref{fig:pwrap_collapse}. As expected, Fortuin-Kasteleyn clusters of the $\pm1$ spins exhibit the standard scaling collapse associated with critical percolation along the continuous transition line. The vacancy sector displays qualitatively different behavior: for $\Delta < \Delta_{\rm tp}$, the apparent wrapping signals are purely finite-size effects and do not show scale-invariant collapse. By contrast, at $\Delta = \Delta_{\rm tp}$ the vacancy wrapping probability exhibits a clear crossing and scaling collapse, demonstrating the onset of correlated critical percolation. This establishes that vacancy fluctuations become scale invariant only at tricriticality, where they introduce an additional collective slow mode absent along the Ising-like critical line. The emergence of this geometrically critical vacancy sector provides direct evidence for the mechanism underlying the abrupt dynamic crossover observed in Fig.~\ref{fig:phd_zexp}(b).

A natural interpretation of this dynamic crossover emerges from the renormalization-group structure of tricriticality. Along the ordinary Ising critical line, the dominant slow fluctuations are carried by the magnetization field, which is directly targeted by nonlocal cluster updates, thereby suppressing critical slowing down. At tricriticality, however, an additional relevant scaling field associated with vacancy-density fluctuations becomes critical. Since standard cluster constructions act only on the $\pm 1$ sector, they accelerate only part of the critical dynamics, while the vacancy sector remains governed by local diffusive relaxation. The onset of correlated vacancy percolation provides the geometric manifestation of this additional slow mode. From this perspective, the observed restoration of local-update scaling reflects the inability of existing hybrid cluster algorithms to couple efficiently to the full tricritical fluctuation spectrum.

\textit{Conclusions.}---In conclusion, we have shown that hybrid cluster-local Monte Carlo dynamics in the two-dimensional Blume-Capel model remain highly efficient along the entire continuous Ising-like transition line but undergo an abrupt loss of acceleration at tricriticality, where the dynamic scaling reverts to that of local Metropolis updates. We identify this crossover with the onset of correlated vacancy percolation, whose emergent system-spanning geometry obstructs the nonlocal relaxation of critical modes. These results reveal a fundamental geometric limitation of cluster acceleration at multicriticality and suggest a broader principle: the efficiency of nonlocal Monte Carlo methods is controlled not only by order-parameter fluctuations but also by whether the update dynamics couple to all relevant critical degrees of freedom. Beyond the Blume-Capel model, this mechanism may also help explain the reduced performance of cluster algorithms in diluted and disordered systems~\cite{ballesteros98}, pointing toward new design principles for efficient simulation strategies in complex many-body systems.

\textit{Acknowledgements.}---A.V. and N.G.F. are grateful to colleagues in the Department of Theoretical Physics at Complutense University of Madrid for their warm hospitality, during which several motivating discussions contributed to this work. N.G.F. would like to thank A. Malakis and F.K. Diakonos for fruitful discussions related to this work. Part of the numerical calculations reported in this paper were performed at the High-Performance Computing cluster CERES of the University of Essex. M.J. and D.-H.K. acknowledge the support from the Regional Innovation System \& Education (RISE) program through the Gwangju RISE center, funded by the Ministry of Education and the Gwangju Metropolitan City (2026-RISE-05-001). The work of A.V. and N.G.F. was supported by the  Engineering and Physical Sciences Research Council (grant EP/X026116/1 is acknowledged). The work of V.M.-M. was partially supported by Ministerio de Ciencia, Innovación y Universidades (Spain) and by the European Regional Development Fund (MCIU/AEI/10.13039/501100011033/FEDER, UE) through grant no. 
 PID2022-136374NB-C21.

 \bibliography{biblio.bib}

@article{blume66,
    title = {{Theory of the First-Order Magnetic Phase Change in U${\mathrm{O}}_{2}$}},
    author = {Blume, M.},
    journal = {Phys. Rev.},
    volume = {141},
    issue = {2},
    pages = {517--524},
    numpages = {0},
    year = {1966},
    month = {1},
    publisher = {American Physical Society},
    doi = {10.1103/PhysRev.141.517},
    url = {https://link.aps.org/doi/10.1103/PhysRev.141.517}
}

@article{capel66a,
    title = {{On the possibility of first-order phase transitions in Ising systems of triplet ions with zero-field splitting}},
    journal = {Physica},
    volume = {32},
    number = {5},
    pages = {966-988},
    year = {1966},
    issn = {0031-8914},
    doi = {https://doi.org/10.1016/0031-8914(66)90027-9},
    url = {https://www.sciencedirect.com/science/article/pii/0031891466900279},
    author = {Capel, H. W.}
}

@article{capel66b,
    title = {{On the possibility of first-order transitions in Ising systems of triplet ions with zero-field splitting II}},
    journal = {Physica},
    volume = {33},
    number = {2},
    pages = {295-331},
    year = {1967},
    issn = {0031-8914},
    doi = {https://doi.org/10.1016/0031-8914(67)90167-X},
    url = {https://www.sciencedirect.com/science/article/pii/003189146790167X},
    author = {Capel, H. W.}
}

@article{capel67,
    title = {{On the possibility of first-order transitions in Ising systems of triplet ions with zero-field splitting III}},
    journal = {Physica},
    volume = {37},
    number = {3},
    pages = {423-441},
    year = {1967},
    issn = {0031-8914},
    doi = {https://doi.org/10.1016/0031-8914(67)90198-X},
    url = {https://www.sciencedirect.com/science/article/pii/003189146790198X},
    author = {Capel, H. W.}
}

@article{zierenberg15,
    title = {{Parallel multicanonical study of the three-dimensional Blume-Capel model}},
    author = {Zierenberg, Johannes and Fytas, Nikolaos G. and Janke, Wolfhard},
    journal = {Phys. Rev. E},
    volume = {91},
    issue = {3},
    pages = {032126},
    numpages = {8},
    year = {2015},
    month = {3},
    publisher = {American Physical Society},
    doi = {10.1103/PhysRevE.91.032126},
    url = {https://link.aps.org/doi/10.1103/PhysRevE.91.032126}
}

@article{kwak15,
    title = {{First-order phase transition and tricritical scaling behavior of the Blume-Capel model: A Wang-Landau sampling approach}},
    author = {Kwak, Wooseop and Jeong, Joohyeok and Lee, Juhee and Kim, Dong-Hee},
    journal = {Phys. Rev. E},
    volume = {92},
    issue = {2},
    pages = {022134},
    numpages = {9},
    year = {2015},
    month = {8},
    publisher = {American Physical Society},
    doi = {10.1103/PhysRevE.92.022134},
    url = {https://link.aps.org/doi/10.1103/PhysRevE.92.022134}
}

@article{zierenberg17,
    title = {{Scaling and universality in the phase diagram of the 2D Blume-Capel model}},
    author = {Zierenberg, Johannes and Fytas, Nikolaos G. and Weigel, Martin and Janke, Wolfhard and Malakis, Anastasios},
    journal = {Eur. Phys. J. Spec. Top.},
    volume = {226},
    issue = {4},
    pages = {789--804},
    numpages = {9},
    year = {2017},
    month = {4},
    issn = {1951-6401},
    doi = {10.1140/epjst/e2016-60337-x},
    url = {https://doi.org/10.1140/epjst/e2016-60337-x}
}

@article{liu23,
  title = {{Critical dynamical behavior of the Ising model}},
  author = {Liu, Zihua and Vatansever, Erol and Barkema, Gerard T. and Fytas, Nikolaos G.},
  journal = {Phys. Rev. E},
  volume = {108},
  issue = {3},
  pages = {034118},
  numpages = {5},
  year = {2023},
  month = {Sep},
  publisher = {American Physical Society},
  doi = {10.1103/PhysRevE.108.034118},
  url = {https://link.aps.org/doi/10.1103/PhysRevE.108.034118}
}

@article{bisson25,
  title = {Universal exotic dynamics in critical mesoscopic systems: Simulating the square root of {A}vogadro's number of spins},
  author = {Bisson, Mauro and Vasilopoulos, Alexandros and Bernaschi, Massimo and Fatica, Massimiliano and Fytas, Nikolaos G. and Pemart\'{\i}n, Isidoro Gonz\'alez-Adalid and Mart\'{\i}n-Mayor, V\'{\i}ctor},
  journal = {Phys. Rev. Res.},
  volume = {7},
  issue = {3},
  pages = {033218},
  numpages = {6},
  year = {2025},
  month = {Sep},
  publisher = {American Physical Society},
  doi = {10.1103/ngkf-7816},
  url = {https://link.aps.org/doi/10.1103/ngkf-7816}
}

@article{fytas18,
    title = {{Universality from disorder in the random-bond Blume-Capel model}},
    author = {Fytas, N. G. and Zierenberg, J. and Theodorakis, P. E. and Weigel, M. and Janke, W. and Malakis, A.},
    journal = {Phys. Rev. E},
    volume = {97},
    issue = {4},
    pages = {040102},
    numpages = {6},
    year = {2018},
    month = {4},
    publisher = {American Physical Society},
    doi = {10.1103/PhysRevE.97.040102},
    url = {https://link.aps.org/doi/10.1103/PhysRevE.97.040102}
}

@book{press92,
    title={{Numerical recipes in C}},
    author={Press, William H and Teukolsky, Saul A. and Vetterling, William T. and Flannery, Brian P. and others},
    year={1992},
    publisher={Cambridge university press Cambridge}
}

@article{akritidis23,
    title = {{Geometric clusters in the overlap of the Ising model}},
    author = {Akritidis, Michail and Fytas, Nikolaos G. and Weigel, Martin},
    journal = {Phys. Rev. E},
    volume = {108},
    issue = {4},
    pages = {044145},
    numpages = {13},
    year = {2023},
    month = {10},
    publisher = {American Physical Society},
    doi = {10.1103/PhysRevE.108.044145},
    url = {https://link.aps.org/doi/10.1103/PhysRevE.108.044145}
}

@book{stauffer_introduction_1994,
    address = {London ; Bristol, PA},
    edition = {Rev., 2nd},
    title = {{Introduction to Percolation Theory}},
    publisher = {Taylor \& Francis},
    author = {Stauffer, Dietrich and Aharony, Amnon},
    year = {1994}
}

@article{swendsen87,
  title = {{Nonuniversal critical dynamics in Monte Carlo simulations}},
  author = {Swendsen, Robert H. and Wang, Jian-Sheng},
  journal = {Phys. Rev. Lett.},
  volume = {58},
  issue = {2},
  pages = {86--88},
  numpages = {0},
  year = {1987},
  month = {Jan},
  publisher = {American Physical Society},
  doi = {10.1103/PhysRevLett.58.86},
  url = {https://link.aps.org/doi/10.1103/PhysRevLett.58.86}
}

@article{wolff89,
  title = {Collective {M}onte {C}arlo Updating for Spin Systems},
  author = {Wolff, Ulli},
  journal = {Phys. Rev. Lett.},
  volume = {62},
  issue = {4},
  pages = {361--364},
  numpages = {0},
  year = {1989},
  month = {Jan},
  publisher = {American Physical Society},
  doi = {10.1103/PhysRevLett.62.361},
  url = {https://link.aps.org/doi/10.1103/PhysRevLett.62.361}
}

@article{FK72a,
    title = {{On the random-cluster model: I. Introduction and relation to other models}},
    journal = {Physica},
    volume = {57},
    number = {4},
    pages = {536-564},
    year = {1972},
    issn = {0031-8914},
    doi = {https://doi.org/10.1016/0031-8914(72)90045-6},
    url = {https://www.sciencedirect.com/science/article/pii/0031891472900456},
    author = {C.M. Fortuin and P.W. Kasteleyn}
}

@article{FK72b,
    title = {{On the random-cluster model II. The percolation model}},
    journal = {Physica},
    volume = {58},
    number = {3},
    pages = {393-418},
    year = {1972},
    issn = {0031-8914},
    doi = {https://doi.org/10.1016/0031-8914(72)90161-9},
    url = {https://www.sciencedirect.com/science/article/pii/0031891472901619},
    author = {C.M. Fortuin}
}

@article{FK72c,
    title = {{On the random-cluster model: III. The simple random-cluster model}},
    journal = {Physica},
    volume = {59},
    number = {4},
    pages = {545-570},
    year = {1972},
    issn = {0031-8914},
    doi = {https://doi.org/10.1016/0031-8914(72)90087-0},
    url = {https://www.sciencedirect.com/science/article/pii/0031891472900870},
    author = {C.M. Fortuin},
}

@incollection{sokal97,
	location = {Boston},
	title = {Monte {C}arlo Methods in Statistical Mechanics: {F}oundations and New Algorithms},
	volume = {361},
	isbn = {978-1-4899-0321-1 978-1-4899-0319-8},
	url = {http://link.springer.com/10.1007/978-1-4899-0319-8_6},
	shorttitle = {Monte Carlo Methods in Statistical Mechanics},
	pages = {131--192},
	booktitle = {Functional Integration},
	publisher = {Springer},
	author = {Sokal, A. D.},
	editor = {{DeWitt}-Morette, Cecile and Cartier, Pierre and Folacci, Antoine},
	urldate = {2021-03-05},
	date = {1997},
	langid = {english},
	doi = {10.1007/978-1-4899-0319-8_6},
	note = {Series Title: {NATO} {ASI} Series},
}

@inbook{janke08, 
    address={Berlin, Heidelberg}, 
    series={{Lecture Notes in Physics}}, title={{Monte Carlo Methods in Classical Statistical Physics}}, volume={739},
    ISBN={978-3-540-74685-0}, 
    url={http://link.springer.com/10.1007/978-3-540-74686-7_4}, 
    doi={10.1007/978-3-540-74686-7_4},
    booktitle={Computational Many-Particle Physics}, 
    publisher={Springer Berlin Heidelberg}, 
    author={Janke, Wolfhard}, 
    editor={Fehske, H. and Schneider, R. and Weiße, A.}, 
    year={2008}, 
    pages={79–140}, 
    collection={Lecture Notes in Physics}
}

@article{ferrenberg:88a,
    author = {Ferrenberg, A. M. and Swendsen, R. H.},
    doi = {10.1103/PhysRevLett.61.2635},
    journal = {Phys. Rev. Lett.},
    pages = {2635--2638},
    title = {New {M}onte {C}arlo technique for studying phase transitions},
    url = {http://dx.doi.org/10.1103/PhysRevLett.61.2635},
    volume = {61},
    year = {1988}
}

@article{li:89,
    author = {Li, Xiao J. and Sokal, Alan D.},
    doi = {10.1103/physrevlett.63.827},
    journal = {Phys. Rev. Lett.},
    number = {8},
    pages = {827--830},
    publisher = {American Physical Society},
    title = {Rigorous lower bound on the dynamic critical exponents of the {Swendsen-Wang} algorithm},
    url = {http://dx.doi.org/10.1103/physrevlett.63.827},
    volume = {63},
    year = {1989}
}

@book{amit_book,
  title={{Field theory, the renormalization group, and critical phenomena: graphs to computers}},
  author={Amit, Daniel J and Martín-Mayor, Víctor},
  year={2005},
  publisher={World Scientific Publishing Company}
}

@article{deng05,
  title={{Percolation between vacancies in the two-dimensional Blume-Capel model}},
  author={Deng, Youjin and Guo, Wenan and Bl{\"o}te, Henk WJ},
  journal={Phys. Rev. E},
  volume={72},
  number={1},
  pages={016101},
  year={2005},
  publisher={APS},
  doi = {10.1103/PhysRevE.72.016101},
  url = {https://journals.aps.org/pre/abstract/10.1103/PhysRevE.72.016101}
}

@article{nightingale96,
  title={{Dynamic exponent of the two-dimensional Ising model and Monte Carlo computation of the subdominant eigenvalue of the stochastic matrix}},
  author={Nightingale, MP and Bl{\"o}te, HWJ},
  journal={Phys. Rev. Lett.},
  volume={76},
  number={24},
  pages={4548},
  year={1996},
  publisher={APS},
  doi = {10.1103/PhysRevLett.76.4548},
  url = {https://journals.aps.org/prl/abstract/10.1103/PhysRevLett.76.4548}
}

@article{jung17,
  title={{First-order transitions and thermodynamic properties in the 2D Blume-Capel model: the transfer-matrix method revisited}},
  author={Jung, Moonjung and Kim, Dong-Hee},
  journal={Eur. Phys. J. B},
  volume={90},
  number={12},
  pages={245},
  year={2017},
  publisher={Springer},
  doi = {10.1140/epjb/e2017-80471-2},
  url = {https://link.springer.com/article/10.1140/epjb/e2017-80471-2}
}

@article{wilding96,
  title={{Tricritical universality in a two-dimensional spin fluid}},
  author={Wilding, NB and Nielaba, P},
  journal={Phys. Rev. E},
  volume={53},
  number={1},
  pages={926},
  year={1996},
  publisher={APS},
  doi = {10.1103/PhysRevE.53.926},
  url = {https://journals.aps.org/pre/abstract/10.1103/PhysRevE.53.926}
}

@article{ziff00,
  title = {{Efficient Monte Carlo Algorithm and High-Precision Results for Percolation}},
  author = {Newman, M. E. J. and Ziff, R. M.},
  journal = {Phys. Rev. Lett.},
  volume = {85},
  issue = {19},
  pages = {4104--4107},
  numpages = {0},
  year = {2000},
  month = {Nov},
  publisher = {American Physical Society},
  doi = {10.1103/PhysRevLett.85.4104},
  url = {https://link.aps.org/doi/10.1103/PhysRevLett.85.4104}
}

@article{mataragkas25a,
  title = {{Tricriticality and finite-size scaling in the triangular Blume-Capel ferromagnet}},
  author = {Mataragkas, Dimitrios and Vasilopoulos, Alexandros and Fytas, Nikolaos G. and Kim, Dong-Hee},
  journal = {Phys. Rev. Res.},
  volume = {7},
  issue = {1},
  pages = {013214},
  numpages = {15},
  year = {2025},
  month = {Feb},
  publisher = {American Physical Society},
  doi = {10.1103/PhysRevResearch.7.013214},
  url = {https://link.aps.org/doi/10.1103/PhysRevResearch.7.013214}
}

@article{mataragkas25b,
  title = {{Transfer-matrix approach to the Blume-Capel model on the triangular lattice}},
  author = {Mataragkas, Dimitrios and Vasilopoulos, Alexandros and Fytas, Nikolaos G. and Kim, Dong-Hee},
  journal = {Phys. Rev. Res.},
  volume = {7},
  issue = {3},
  pages = {033240},
  numpages = {10},
  year = {2025},
  month = {Sep},
  publisher = {American Physical Society},
  doi = {10.1103/jfl3-f4kd},
  url = {https://link.aps.org/doi/10.1103/jfl3-f4kd}
}

@article{moueddene24,
  title = {{Critical and tricritical behavior of the $d=3$ Blume-Capel model: Results from small-scale Monte Carlo simulations}},
  author = {Moueddene, Le\"{\i}la and Fytas, Nikolaos G. and Berche, Bertrand},
  journal = {Phys. Rev. E},
  volume = {110},
  issue = {6},
  pages = {064144},
  numpages = {12},
  year = {2024},
  month = {Dec},
  publisher = {American Physical Society},
  doi = {10.1103/PhysRevE.110.064144},
  url = {https://link.aps.org/doi/10.1103/PhysRevE.110.064144}
}

@article{ballesteros98,
  title = {{Critical exponents of the three-dimensional diluted Ising model}},
  author = {Ballesteros, H. G. and Fern\'andez, L. A. and Mart\'{\i}n-Mayor, V. and Mu\~noz Sudupe, A. and Parisi, G. and Ruiz-Lorenzo, J. J.},
  journal = {Phys. Rev. B},
  volume = {58},
  issue = {5},
  pages = {2740--2747},
  numpages = {0},
  year = {1998},
  month = {Aug},
  publisher = {American Physical Society},
  doi = {10.1103/PhysRevB.58.2740},
  url = {https://link.aps.org/doi/10.1103/PhysRevB.58.2740}
}

@article{fytas-selke13,
  title={{Wetting and interfacial adsorption in the Blume-Capel model on the square lattice}},
  author={Fytas, N. G. and Selke, W.},
  journal={Eur. Phys. J. B},
  volume={86},
  pages={1--7},
  year={2013},
  publisher={Springer},
  doi = {10.1140/epjb/e2013-40475-6},
  url = {https://link.springer.com/article/10.1140/epjb/e2013-40475-6}
}

@article{salas00,
  title={{Universal amplitude ratios in the critical two-dimensional Ising model on a torus}},
  author={Salas, Jes{\'u}s and Sokal, Alan D},
  journal={J. Stat. Phys.},
  volume={98},
  number={3},
  pages={551--588},
  year={2000},
  publisher={Springer},
  doi = {https://doi.org/10.1023/A:1018611122166},
  url = {https://link.springer.com/article/10.1023/A:1018611122166#citeas}
}

@article{caselle:02,
doi = {10.1088/0305-4470/35/23/305},
url = {https://dx.doi.org/10.1088/0305-4470/35/23/305},
year = {2002},
month = {may},
publisher = {},
volume = {35},
number = {23},
pages = {4861},
author = {Caselle, Michele and Hasenbusch, Martin and  Pelissetto, Andrea and  Vicari, Ettore},
title = {{Irrelevant operators in the two-dimensional Ising model}},
journal = {J. Phys. A: Math. Gen.},
}

@article{silva06,
  title = {{Wang-Landau Monte Carlo simulation of the Blume-Capel model}},
  author = {Silva, C. J. and Caparica, A. A. and Plascak, J. A.},
  journal = {Phys. Rev. E},
  volume = {73},
  issue = {3},
  pages = {036702},
  numpages = {6},
  year = {2006},
  month = {Mar},
  publisher = {American Physical Society},
  doi = {10.1103/PhysRevE.73.036702},
  url = {https://link.aps.org/doi/10.1103/PhysRevE.73.036702}
}

@article{toldin17,
  title = {{Critical behavior in the presence of an order-parameter pinning field}},
  author = {Parisen Toldin, Francesco and Assaad, Fakher F. and Wessel, Stefan},
  journal = {Phys. Rev. B},
  volume = {95},
  issue = {1},
  pages = {014401},
  numpages = {20},
  year = {2017},
  month = {Jan},
  publisher = {American Physical Society},
  doi = {10.1103/PhysRevB.95.014401},
  url = {https://link.aps.org/doi/10.1103/PhysRevB.95.014401}
}

@article{hasenbusch20,
  title = {{Dynamic critical exponent $z$ of the three-dimensional Ising universality class: Monte Carlo simulations of the improved Blume-Capel model}},
  author = {Hasenbusch, Martin},
  journal = {Phys. Rev. E},
  volume = {101},
  issue = {2},
  pages = {022126},
  numpages = {14},
  year = {2020},
  month = {Feb},
  publisher = {American Physical Society},
  doi = {10.1103/PhysRevE.101.022126},
  url = {https://link.aps.org/doi/10.1103/PhysRevE.101.022126}
}

@article{binder81,
  title = {{Critical Properties from Monte Carlo Coarse Graining and Renormalization}},
  author = {Binder, K.},
  journal = {Phys. Rev. Lett.},
  volume = {47},
  issue = {9},
  pages = {693--696},
  numpages = {0},
  year = {1981},
  month = {Aug},
  publisher = {American Physical Society},
  doi = {10.1103/PhysRevLett.47.693},
  url = {https://link.aps.org/doi/10.1103/PhysRevLett.47.693}
}

@article{qian05,
  title = {{Dilute Potts model in two dimensions}},
  author = {Qian, Xiaofeng and Deng, Youjin and Bl\"ote, Henk W. J.},
  journal = {Phys. Rev. E},
  volume = {72},
  issue = {5},
  pages = {056132},
  numpages = {15},
  year = {2005},
  month = {Nov},
  publisher = {American Physical Society},
  doi = {10.1103/PhysRevE.72.056132},
  url = {https://link.aps.org/doi/10.1103/PhysRevE.72.056132}
}

@article{blote19,
  title = {Revisiting the field-driven edge transition of the tricritical two-dimensional Blume-Capel model},
  author = {Bl\"ote, Henk W. J. and Deng, Youjin},
  journal = {Phys. Rev. E},
  volume = {99},
  issue = {6},
  pages = {062133},
  numpages = {4},
  year = {2019},
  month = {Jun},
  publisher = {American Physical Society},
  doi = {10.1103/PhysRevE.99.062133},
  url = {https://link.aps.org/doi/10.1103/PhysRevE.99.062133}
}

\appendix

\section{End Matter}

\textit{Appendix A: Critical-point estimates.}---We summarize here high-precision estimates of the critical points along the continuous transition line of the square-lattice Blume-Capel model, including the benchmark case $\Delta = 0$~\cite{bisson25}, with inverse critical temperatures determined to a precision of order $10^{-7}$.
Equilibrium simulations were performed at $\Delta = 0$, $1$, $1.5$, and $1.8$ on square lattices of linear size $L=2^n$ ($3\le n\le12$) with periodic boundary conditions. Configurations were generated using a hybrid Swendsen-Wang--Metropolis scheme consisting of $10$ Metropolis sweeps followed by one Swendsen-Wang update restricted to the $\pm1$ sector. For each lattice size, $100$ independent runs of $4\times10^5$ measurements were collected and analyzed via histogram reweighting~\cite{ferrenberg:88a}.

Critical-point estimates were obtained from the Binder cumulant,
$U_4=\langle m^4\rangle/\langle m^2\rangle^2$~\cite{binder81}, and the second-moment correlation-length ratio $\xi/L$, with
$\xi= \frac{1}{2\sin(\pi/L)}
\sqrt{\frac{\chi(\mathbf{0})}{\chi(\mathbf{k}_{\rm min})}-1}$~\cite{amit_book}.
For the two-dimensional Ising universality class on the square lattice, the corresponding universal values are known to high precision:
$U_4^{\ast}=1.1679229(47)$ and $(\xi/L)^{\ast}=0.9050488292(4)$~\cite{salas00}. Since the Blume-Capel model belongs to this universality class throughout the continuous transition regime, pseudo-critical inverse temperatures $\beta^{\ast}_{L,g}$ ($g\in\{U_4,\xi/L\}$) were determined from the conditions $U_4(L,\beta_{L,U_4})=U_4^{\ast}$ and $\xi(L,\beta_{L,\xi/L})/L=(\xi/L)^{\ast}$.

\begin{figure}[ht!]
    \centering
    \includegraphics[width=0.85\linewidth]{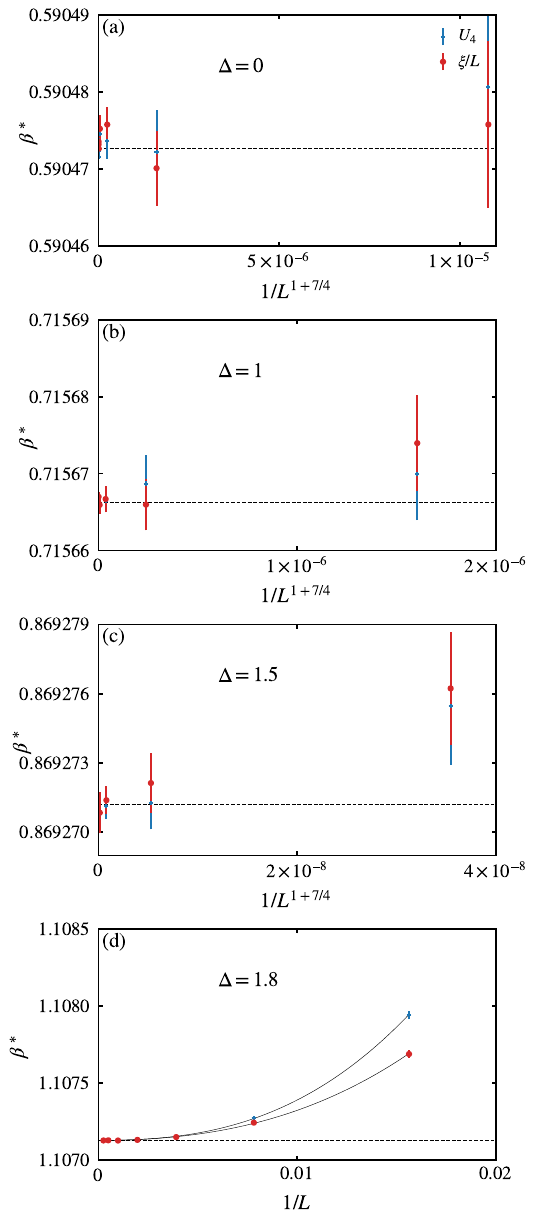}
    \caption{Finite-size scaling of pseudo-critical inverse temperatures $\beta^{\ast}$ obtained from the Binder cumulant ($U_4$) and correlation-length ratio ($\xi/L$) for the values of $\Delta$ indicated in the panels. Joint fits yield the thermodynamic critical inverse temperatures $\beta_{\rm c}$, marked by the horizontal dashed lines. While constant extrapolations suffice for $\Delta \leq 1.5$, explicit scaling corrections are required at $\Delta=1.8$ due to proximity to tricriticality.}
    \label{fig:beta_c}
\end{figure}

These estimates were extrapolated according to
\[
\beta^{\ast}_{L,g}=\beta_{\rm c}+\mathcal{A}_gL^{-(y_t+\omega)}+\cdots,
\]
with thermal exponent $y_t=1$ for the Ising universality class~\cite{bisson25,mataragkas25a} and $\mathcal{A}$ nonuniversal fitting constants. For $\Delta=0$, $1$, and $1.5$, the pseudo-critical temperatures converge rapidly and are accurately described by constant joint fits over sufficiently large system sizes, consistent with Ref.~\cite{bisson25}. For $\Delta=1.8$, close to the tricritical point $\Delta_{\rm tp}=1.96582(1)$~\cite{qian05,kwak15,jung17,blote19}, stable extrapolations required explicit correction terms associated with both the leading Ising analytic exponent $\omega=7/4$~\cite{caselle:02} and tricritical crossover corrections, which we parametrize using the tangent tricritical scaling field with exponent $y_g^{(\rm tp)}=4/5$~\cite{mataragkas25a}. Fits based on alternative tricritical correction forms yield statistically consistent estimates. The resulting extrapolations are shown in Fig.~\ref{fig:beta_c}, with final estimates summarized in Table~\ref{tab:beta_c}.

\begin{table}[ht!]
\centering
\caption{Critical inverse temperatures $\beta_{\rm c}$ of the square-lattice Blume-Capel model along the continuous transition line, obtained from joint finite-size scaling fits to Binder-cumulant and correlation-length-ratio universal crossing points. Also shown are the minimum system size included in each fit, $L_{\rm min}$, and the corresponding goodness-of-fit statistics, $\chi^2/{\rm DOF}$, where DOF denotes the number of degrees of freedom~\cite{press92}.}
\begin{tabular*}{1\columnwidth}{@{\extracolsep{\fill}} lccc @{}}
\hline\hline
$\Delta$ & \multicolumn{1}{c}{$\beta_{\rm c}$} & $L_{\rm min}$ & \multicolumn{1}{c}{$\chi^2/{\rm DOF}$}  \\ \hline
$0$   & $0.5904727(2)$ & 64  & 1.69  \\ \hline
$1$   & $0.7156663(2)$ & 128 & 0.37 \\ \hline
$1.5$ & $0.8692713(3)$ & 512 & 1.17  \\ \hline
$1.8$ & $1.1071271(2)$ & 64  & 0.51 \\
\hline\hline
\end{tabular*}\label{tab:beta_c}
\end{table}

\textit{Appendix B: Definition of Monte Carlo time.}---To enable a quantitative comparison between different update protocols, we define the Monte Carlo time as
\[
\tau = \tau_{\rm step}\left[
n_{\rm M}
+\frac{\langle N_c^{\rm W}\rangle}{N}n_{\rm W}
+\frac{\langle E_\Delta\rangle}{N}n_{\rm SW}
\right],
\]
where $\tau_{\rm step}$ denotes the number of measurement intervals, each consisting of $n_{\rm M}$ Metropolis sweeps, $n_{\rm W}$ Wolff updates, and $n_{\rm SW}$ Swendsen-Wang updates. The factors $\langle E_\Delta\rangle/N$ and $\langle N_c^{\rm W}\rangle/N$, with $\langle N_c^{\rm W}\rangle$ the average Wolff-cluster size, rescale cluster updates into equivalent lattice-sweep units, thereby defining a common time scale for all hybrid protocols. In Fig.~\ref{fig:phd_zexp}(b), the update schedules correspond to $(n_{\rm M},n_{\rm W},n_{\rm SW})=(1,0,0)$ for Metropolis, $(1,1,0)$ for Wolff$+$Metropolis, and $(1,0,1)$ for Swendsen-Wang$+$Metropolis.

\textit{Appendix C: Dynamic critical exponent.}---For the benefit of the reader, the numerical estimates of the dynamic critical exponent $z_{\rm exp}$ shown in Fig.~\ref{fig:phd_zexp}(b) are listed in Table~\ref{tab:zexp}. The data quantify the persistence of efficient cluster dynamics along the critical line and their breakdown at tricriticality.

\begin{table}[ht!]
\caption{Numerical estimates of the dynamic critical exponent $z_{\rm exp}$ along the continuous transition line of the square-lattice Blume-Capel model, corresponding to Fig.~\ref{fig:phd_zexp}(b) of the main text.}
\begin{tabular*}{0.99\columnwidth}{@{\extracolsep{\fill}} lccc @{}}
\hline\hline
$\Delta$ &  & $z_{\rm exp}$ &      \\
\hline
& M & W$+$M & SW$+$M\\
\hline
$0$   & $2.165(6)$ & $0.368(4)$  & $0.248(7)$ \\
\hline
$1$   & $2.154(9)$  & $0.336(4)$ &  $0.230(2)$   \\
\hline
$1.5$ & $2.157(7)$ & $0.284(2)$  & $0.224(4)$\\
\hline
$1.8$ & $2.124(6)$ &  $0.235(6)$ & $0.214(6)$ \\ 
\hline \hline
\multicolumn{4}{c}{\textbf{Tricritical Point}}      \\
 \hline
$1.96582$ & $2.123(6)$  & $1.929(4)$ & $1.942(5)$ \\
\hline \hline
\end{tabular*}
\label{tab:zexp}
\end{table}

\begin{figure}[ht!]
    \centering
    \includegraphics[width=0.85\linewidth]{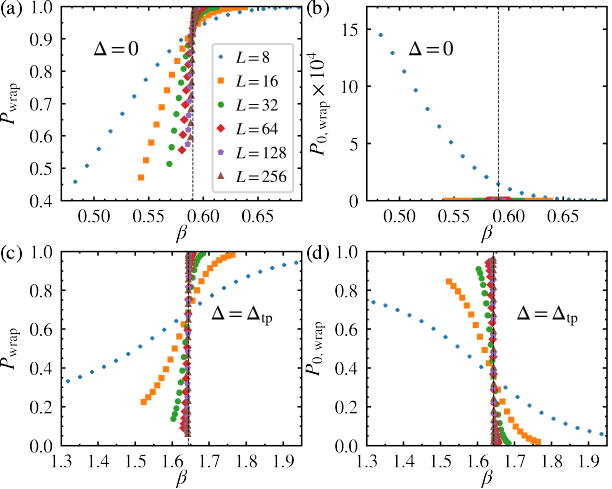}
    \caption{The unrescaled version of wrapping probabilities of Fortuin-Kasteleyn clusters of $\pm 1$ spins ($P_{\rm wrap}$, left column) and geometric vacancy clusters ($P_{0,{\rm wrap}}$, right column) as functions of the inverse temperature $\beta$. The vertical line in each panel indicates the transition point.}
    \label{fig:pwrap}
\end{figure}

\textit{Appendix D: Wrapping probabilities.}---For completeness, we include in Fig.~\ref{fig:pwrap} the unrescaled finite-size data corresponding to the scaling plots shown in Fig.~\ref{fig:pwrap_collapse} of the main text.

\end{document}